\RequirePackage{fix-cm}
\documentclass[smallextended]{svjour3}       
\smartqed  
\usepackage{graphicx}
\begin{document}

\title{Cosmic bulk viscosity through backreaction}


\author{Rodrigo M. Barbosa \and   \\Eddy~G.~Chirinos I. \and \\Winfried Zimdahl    \and\\
        Oliver~F.~Piattella
}


\institute{Rodrigo M. Barbosa \at
              Universidade Federal do Esp\'{\i}rito Santo,
Departamento
de F\'{\i}sica,
Av. Fernando Ferrari, 514, Campus de Goiabeiras, CEP 29075-910,
Vit\'oria, Esp\'{\i}rito Santo, Brazil \\
           \and
           Eddy~G.~Chirinos I. \at
              Universidade Federal do Esp\'{\i}rito Santo,
Departamento
de F\'{\i}sica,
Av. Fernando Ferrari, 514, Campus de Goiabeiras, CEP 29075-910,
Vit\'oria, Esp\'{\i}rito Santo, Brazil \\
           \and
           Winfried Zimdahl\at
              Universidade Federal do Esp\'{\i}rito Santo,
Departamento
de F\'{\i}sica,
Av. Fernando Ferrari, 514, Campus de Goiabeiras, CEP 29075-910,
Vit\'oria, Esp\'{\i}rito Santo, Brazil \\
              \email{winfried.zimdahl@pq.cnpq.br}           
           \and
           Oliver F. Piattella \at
              Universidade Federal do Esp\'{\i}rito Santo,
Departamento
de F\'{\i}sica,
Av. Fernando Ferrari, 514, Campus de Goiabeiras, CEP 29075-910,
Vit\'oria, Esp\'{\i}rito Santo, Brazil \\
              \email{oliver.piattella@pq.cnpq.br}           
}

\date{Received: date / Accepted: date}

\maketitle

\begin{abstract}
We consider an effective viscous pressure as the result of a backreaction of inhomogeneities within Buchert's formalism. The use of an effective metric with a time-dependent curvature radius allows us to calculate the luminosity distance of the backreaction model. This quantity is different from its counterpart for a
``conventional" spatially flat bulk viscous fluid universe.
Both expressions are tested against the SNIa data of the Union2.1 sample with only marginally different results
for the distance-redshift relation and in accordance with the $\Lambda$CDM model. Future observations are expected to be able to discriminate among these models on the basis of indirect measurements of the curvature evolution.
\end{abstract}

\section{Introduction}
Notwithstanding the tremendous success of the cosmological standard model in fitting an impressive amount of data
(for a recent summary of significant tensions, however, see \cite{buchert15}), the physical nature of some of its main ingredients, notably the components of the ``dark sector" remains vague.
Despite of many efforts in numerous ongoing projects no direct detection of dark matter has been realized so far. Dark energy is even more elusive unless one is willing to accept a cosmological constant as an additional constant of nature.
There is therefore continuing motivation to keep an eye on approaches that modify or alter the standard picture.
A rather radical route is to abandon General Relativity (GR) as the correct theory of gravitation and to replace it by a more general framework in which the dark components are no longer matter components but become part of the geometric
sector of the generalized field equations (see, e.g., \cite{copeland,trodden2015}).
A more conservative way which also tries to understand the dark sector of the standard model in geometrical terms
is the backreaction approach. This approach generalizes the cosmological principle insofar as it does not take for granted the existence of a highly symmetric homogeneous and isotropic background but regards the Universe to be only statistically homogeneous and isotropic. The problem that arises is how to perform the required averages over  inhomogeneous matter and space-time configurations.
This is intimately related to the ``fitting problem" \cite{EllisStoeger:fitting 1987}, a fundamental issue for theoretical and observational cosmology.

While the appropriate definition of averages in GR and particularly in cosmology remains an open issue, there are approaches which are supposed to capture important aspects of the problem.
Several averaging procedures have been discussed in the literature \cite{Zal1:1992,Zal2:1993,Buchert:1999er,Sussman08,Sussman11,Kor:2010,Skarke}.
Reviews of the status of the field can be found in \cite{EllisCQG,WiltshireCQG,BuchertCQG,RasaCQG,KolbCQG}.

Taking into account backreaction from spatial inhomogeneities on the average large-scale homogeneous cosmological dynamics has been considered to be a promising route to describe the evolution of the late, matter-dominated Universe. Here we focus on Buchert's approach which relies of an average of scalar quantities over spatial hypersurfaces \cite{Buchert:1999er}.
In Buchert's equations for the volume scale factor of an irrotational dust universe there appear two types of additional contributions compared with the standard Friedmann-Lema\^{\i}tre-Robertson-Walker (FLRW) dynamics: a kinematic backreaction and an averaged curvature term. These two terms are related to each other by a consistency condition.
It is a specific feature of Buchert's approach that kinematic backreaction and averaged curvature can be combined into an effective energy density and an effective pressure. Then the dynamics of the volume scale factor formally coincides with that of a two-fluid universe, one component being dust, the other one the effective backreaction fluid. The difference is that the volume scale factor is not the scale factor of a Robertson-Walker (RW) metric since,
through the averaging operation, the scale factor must be scale-dependent.

The backreaction fluid has been modeled as a scalar field \cite{Buchert:2006ya} and as a Chaplygin gas \cite{RoyBuchert:2010}.
In conventional standard cosmology a scalar field is introduced as an additional component in the matter sector which is supposed to describe dark energy, a component with a sufficiently large negative pressure to account for the accelerated expansion of the Universe.
In the backreaction picture, on the other hand, it is not an additional component but it is of geometric origin in an otherwise pure dust universe. A similar comment holds for the Chaplygin gas.
Whether or not the backreaction is sufficiently strong to reproduce the observational data without the necessity of  dark-energy is a much debated problem \cite{GreenWald,AntiWald}.

While conceptually appealing, Buchert's backreaction picture so far lacks a natural method to describe light propagation and hence relate the general dynamics to observations which require data on the light cone, not on the spatial hypersurfaces on which the averages are performed.
This raises the problem of how to relate the quantities of the backreaction formalism to observations.
There is no space-time metric to which the usual condition $ds^{2} =0$ for light propagation could be applied.
Studies of light propagation in backreaction cosmology have been performed, e.g., in \cite{rasa2009,rasa2010} and \cite{schwarz2014}.
A provisional way to deal with this issue has been to assume the existence of a template metric \cite{ParanjapeSingh:2006,Larena:2008be} in which the volume scale factor is supposed to play the role of a ``conventional" scale factor as in the RW metric, although the template metric is not required to be a solution of the field equations.
Moreover, the averaged curvature is assumed to be describable by a curvature term in the template metric.
Since, however, there is, in general, no constant curvature as in FLRW models, one has a different curvature ``constant" on each slice $t=\ $ constant.
Through this procedure the curvature evolution is made compatible with the exact
average properties of the universe model.

Obviously one has to face the question, whether an effective backreaction fluid (or scalar field) of geometric origin is observationally distinguishable from a ``true" fluid component, i.e., an additional form of matter as, e.g., quintessence.
We shall address this point here using a bulk viscous fluid as an example.
Like Chaplygin gases, bulk viscous fluids have been discussed as potential candidates for a unified description of the dark sector \cite{rose,Szydlowski,BVM,avelino,barrow,avelino10,VDF,VDF2}. Bulk viscous models can account for the background dynamics and for the large-scale-structure data \cite{VDF2}. They do less well in describing the CMB spectrum \cite{barrow}.
The idea of a viscosity-dominated late
epoch of the Universe with accelerated expansion was already mentioned in \cite{PadChitre:1987}, long before
the direct observational evidence for an accelerated expansion through the SN Ia data.

Assuming the dark sector of the cosmic medium to behave as a bulk viscous fluid has always provoked the question of the origin of the viscosity. Nonstandard interactions have to be postulated to generate a negative pressure of the required order
\cite{antifr:2001,NJP:2003}. This reflects our ignorance concerning the nature of the dark sector.
If, however, a cosmic bulk viscosity  can arise as the result of a backreaction of inhomogeneities, it would have a natural geometric interpretation.
In this new context also the observational status of the model has to be reconsidered.

 Our strategy is to start with a pure dust universe and to identify the kinematic backreaction and the averaged curvature that would dynamically be equivalent to a universe made of nonrelativistic matter and a bulk viscous fluid.
To study observational implications of this approach we calculate the luminosity distance in the backreaction model under the additional assumption of the existence of an effective homogeneous metric in which the curvature ``constant" is different for each slice $t=\ $ constant.
Using the Union2.1 sample we compare this result with the luminosity distance of a ``conventional" bulk viscous fluid dynamics in a spatially flat universe.
We regard this to be a simple analytically tractable toy model to demonstrate various aspects of cosmological backreaction dynamics.

The paper is organized as follows.
Section \ref{buchert} recalls the basic relations of Buchert's backreaction formalism.
The representation in terms of an effective fluid is the subject of section \ref{fluid}.
In section \ref{vfluid} we introduce a domain-dependent effective viscous pressure which will be related
to effective curvature and kinematical backreaction in section \ref{vback}.
With the help of an effective metric we determine the luminosity distance within this model in
section \ref{effective} and discuss observational consequences. Finally, section \ref{summary} summarizes
our main conclusions.

\section{The Buchert equations}
\label{buchert}

Assuming a matter content of irrotational dust, the Buchert equations are \cite{Buchert:1999er}
\begin{equation}\label{Ftype}
\left(\frac{\dot{a}_{D}}{a_{D}}\right)^{2}
- \frac{8\pi G}{3}\left\langle\rho_{m}\right\rangle_{D} = - \frac{\mathcal{R}_{D} + \mathcal{Q}_{D}}{6},
\end{equation}
\begin{equation}\label{ddotb}
\frac{\ddot{a}_{D}}{a_{D}} + \frac{4\pi G}{3}\left\langle\rho_{m}\right\rangle_{D} = \frac{\mathcal{Q}_{D}}{3}\
\end{equation}
and
\begin{equation}\label{mbalD}
\left\langle\rho_{m}\right\rangle_{D}^{\displaystyle\cdot} + 3 \frac{\dot{a}_{D}}{a_{D}}\left\langle\rho_{m}\right\rangle_{D} = 0\ ,
\end{equation}
where $\mathcal{Q}_{D}$ is the kinematical backreaction
\begin{equation}\label{QD}
\mathcal{Q}_{D} = \frac{2}{3}\left\langle\left(\Theta - \left\langle\Theta\right\rangle_{D}\right)^{2}\right\rangle_{D}
- 2 \left\langle\sigma^{2}\right\rangle_{D}\
\end{equation}
and $\mathcal{R}_{D}$ is the averaged three curvature
\begin{equation}\label{RD}
\mathcal{R}_{D} =\left\langle ^{3}R\right\rangle_{D},
\end{equation}
obtained from the three-curvature scalar $^{3}R$ of the hypersurface $t= $ constant.
The quantities $\mathcal{Q}_{D}$ and $\mathcal{R}_{D}$ obey the consistency relation
\begin{equation}\label{intcond}
\frac{1}{a_{D}^{6}}\left(\mathcal{Q}_{D}a_{D}^{6}\right)^{\displaystyle\cdot}
+ \frac{1}{a_{D}^{2}}\left(\mathcal{R}_{D}a_{D}^{2}\right)^{\displaystyle\cdot} = 0.
\end{equation}
The averages that appear in these equations are volume averages of scalar quantities $S(t,r)$ over a rest mass preserving domain $D$ in the $t$ = const hypersurfaces,
\begin{equation}\label{avSgen}
\left\langle S\right\rangle_{D} = \frac{1}{V_{D}}\int_{D}S(t,r)\sqrt{|g_{ij}|}d^{3}r , \qquad V_{D} = \int_{D}\sqrt{|g_{ij}|}d^{3}r ,
\end{equation}
where $|g_{ij}|$ is the determinant of the spatial three-metric on time-orthogonal hypersurfaces. The volume scale factor is defined by
\begin{equation}\label{aD}
a_{D}(t) = \left[\frac{V_{D}(t)}{V_{D0}}\right]^{1/3} ,
\end{equation}
where $V_{D0}= V_{D}(t_{0})$ is a reference volume of the domain at a time $t_{0}$ which will be identified with the present time. Throughout we assume a nonsingular evolution of the dust configuration, something which is not
necessarily guaranteed.

Introducing the parameters
\begin{equation}\label{defOmega}
\mathcal{H_{D}} = \frac{\dot{a}_{D}}{a_{D}},\quad \Omega_{m}^{D} = \frac{8\pi G}{3\mathcal{H_{D}}^{2}}\left\langle\rho_{m}\right\rangle_{D},
\quad
\Omega_{Q}^{D} = - \frac{\mathcal{Q}_{D}}{6\mathcal{H_{D}}^{2}}, \quad
\Omega_{\mathcal{R}}^{D} = - \frac{\mathcal{R}_{D}}{6\mathcal{H_{D}}^{2}},
\end{equation}
the Friedmann-type equation (\ref{Ftype}) is written as
\begin{equation}\label{sum}
\Omega_{m}^{D} + \Omega_{Q}^{D} + \Omega_{R}^{D} = 1.
\end{equation}
Additionally to the matter contribution $\Omega_{m}^{D}$, there appear the fractional quantities
$\Omega_{Q}^{D}$ and $\Omega_{R}^{D}$ which quantify the impact of the kinematical backreaction and of the averaged spatial curvature, respectively, on the cosmological dynamics.

\section{Effective fluid description}
\label{fluid}
\subsection{General relations}
We may define an effective backreaction fluid (see \cite{buchert07}) by (the subindex b denotes backreaction)
\begin{equation}\label{rhob}
\rho_{bD} = - \frac{1}{16\pi G}\left(\mathcal{Q}_{D} + \mathcal{R}_{D}\right)\ , \qquad p_{bD} = - \frac{1}{16\pi G}\left(\mathcal{Q}_{D} - \frac{\mathcal{R}_{D}}{3}\right),
\end{equation}
where $\rho_{bD}$ is an effective energy density and $p_{bD}$ is an effective pressure.
This leads to the Friedmann-type set of equations
\begin{equation}\label{frb}
\left(\frac{\dot{a}_{D}}{a_{D}}\right)^{2}
- \frac{8\pi G}{3}\left(\left\langle\rho_{m}\right\rangle_{D} + \rho_{bD}\right) = 0
\end{equation}
and
\begin{equation}\label{ddotb2}
\frac{\ddot{a}_{D}}{a_{D}} + \frac{4\pi G}{3}\left(\left\langle\rho_{m}\right\rangle_{D} + \rho_{bD} + 3 p_{bD}\right) = 0,
\end{equation}
which implies
\begin{equation}\label{consb}
\dot{\rho}_{bD} + 3 \frac{\dot{a}_{D}}{a_{D}}\left(\rho_{bD} + p_{bD}\right) = 0 \
\end{equation}
for the backreaction fluid.
Then one may introduce a total energy density $\rho_{D}$,
\begin{equation}\label{rhoT}
\rho_{D} = \left\langle\rho_{m}\right\rangle_{D} + \rho_{bD},
\end{equation}
together with the total pressure $p_{D} \equiv p_{bD}$ such that the conservation law
\begin{equation}\label{consrhoT}
\dot{\rho}_{D} + 3 \frac{\dot{a}_{D}}{a_{D}}\left(\rho_{D} + p_{D}\right) = 0
\end{equation}
holds. The effective equation-of-state (EoS) parameter of the backreaction fluid is
\begin{equation}\label{EoS}
\frac{p_{bD}}{\rho_{bD}} = \frac{\mathcal{Q}_{D} - \frac{1}{3}\mathcal{R}_{D}}{\mathcal{Q}_{D} + \mathcal{R}_{D}}.
\end{equation}
A domain dependent deceleration parameter is defined by
\begin{equation}\label{qD}
q_{D} \equiv - \frac{\ddot{a}_{D}a_{D}}{\dot{a}_{D}^{2}},
\end{equation}
which in terms of $\Omega_{Q}^{D}$  and $\Omega_{Q}^{D}$ can be written as
\begin{equation}\label{qDO}
q_{D} = \frac{1}{2} + \frac{3}{2}\Omega_{Q}^{D}\left[1 - \frac{1}{3}\frac{\Omega_{R}^{D}}{\Omega_{Q}^{D}}\right].
\end{equation}
The effective fluid picture has been used to model the cosmological scalar-field dynamics \cite{Buchert:2006ya} and the unified description of the dark sector with the help of a Chaplygin gas \cite{RoyBuchert:2010} in terms of  backreaction and averaged curvature variables.

\subsection{Example cosmological constant}

The simplest example to be constructed is that of a cosmological constant described through effective fluid quantities.
 A dynamics that mimics a cosmological constant would be equivalent to $p_{b}^{D} = - \rho_{b}^{D}$.
 Combining this condition with relations (\ref{rhob}) we find
\begin{equation}\label{}
p_{bD} = -  \rho_{bD} \quad \rightarrow \quad \mathcal{R}_{D} = - 3 {\cal{Q}}_{D}\ .
\end{equation}
The resulting explicit expressions for $p_{bD}$ and $\rho_{bD}$ then are
\begin{equation}\label{}
p_{bD} = - \frac{1}{8\pi G} {\cal{Q}}_{D}\quad \mathrm{and} \quad \rho_{bD} = \frac{1}{8\pi G} {\cal{Q}}_{D}
\ .
\end{equation}
A domain-dependent constant kinematical backreaction ${\cal{Q}}_{D}$ together with a constant negative averaged curvature $\mathcal{R}_{D} = - 3 {\cal{Q}}_{D}$ produces the same dynamics as a (domain-dependent) cosmological constant $\Omega_{\Lambda}^{D}$,
\begin{equation}\label{LambdaD}
\Omega_{\Lambda}^{D} = \frac{1}{3}\frac{\mathcal{Q}_{D}}{\mathcal{H}_{D0}^{2}} = - \frac{1}{9}\frac{\mathcal{R}_{D}}{\mathcal{\mathcal{H}}_{D0}^{2}}.
\end{equation}
Notice that a positive $\mathcal{Q}_{D}$ corresponds to a negative $\Omega_{Q}^{D}$ and a negative averaged curvature $\mathcal{R}_{D}$ with $\Omega_{R}^{D} > 0$ and a curvature radius
\begin{equation}\label{RcDL}
\mathcal{R}_{cD}^{\Lambda} = \frac{c}{a_{D}}\sqrt{\frac{-6}{\mathcal{R}_{D}}}
= \frac{c}{a_{D}}\sqrt{\frac{2}{\mathcal{Q}}_{D}}\ .
\end{equation}
Its present value
\begin{equation}\label{RcDL0}
\mathcal{R}_{cD0}^{\Lambda} = \frac{c}{\mathcal{H}_{D0}}\sqrt{\frac{2}{3\Omega_{\Lambda}^{D}}}
\end{equation}
is of the order of the Hubble length of the domain under consideration. If we take this domain to be the observable Universe, this curvature radius is of the gigaparsec scale.
Can one discriminate between an effective cosmological constant as a result of backreaction and a ``true" cosmological constant? We shall come back to this in a later section.

\section{Domain-dependent viscous fluid}
\label{vfluid}

Bulk viscous fluids have been discussed as potential candidates for a unified description of dark matter and dark energy \cite{rose,Szydlowski,BVM,avelino,barrow,avelino10,VDF,VDF2}. Our interest here is to reconsider such models in a backreaction context.

A bulk viscous fluid in the domain $D$ is characterized by an EoS
\begin{equation}
 p_{vD} = p_{D} = - \zeta_{D} \left\langle\Theta\right\rangle_{D} = - 3\zeta_{D} \frac{\dot{a}_{D}}{a_{D}}\ .
\label{pvD}
\end{equation}
We have replaced here the previously introduced general backreaction index b by a subindex v which stands for ``viscous" to indicate that we are dealing with the special case of a viscous fluid.
For simplicity we shall assume that $\zeta_{D}$ remains constant over the domain $D$.

With $\rho_{D}^{\prime} \equiv d \rho_{D}/d a_{D}$ we may write (\ref{consrhoT})
with (\ref{pvD}) as
\begin{equation}
\rho_{D}^{\prime} = - \frac{3}{a_{D}}\left(\rho_{D} - 3 \zeta_{D}\frac{\dot{a}_{D}}{a_{D}}\right)
= - \frac{3}{a_{D}}\left(\rho_{D} - 3 \zeta_{D}\sqrt{\frac{8\pi G}{3}}\sqrt{\rho_{D}}\right)\ ,
\end{equation}
where we have used (\ref{frb}).
Integration yields
\begin{equation}
\rho_{D} = \left[A_{D} + \left(\sqrt{\rho_{D0}} - A_{D}\right)a_{D}^{-3/2}\right]^{2}\qquad\qquad\qquad\qquad\qquad\qquad \nonumber
\end{equation}
\begin{equation}
\label{rhoD}
\qquad\qquad\quad \Rightarrow\quad
\mathcal{H}_{D} = \sqrt{\frac{8\pi G}{3}}\left[A_{D} + \left(\sqrt{\rho_{D0}} - A_{D}\right)a_{D}^{-3/2}\right],
\end{equation}
where $\rho_{D0}$ is the present value of $\rho_{D}$ and
\begin{equation}\label{defA}
A_{D} \equiv 3\zeta_{D}\sqrt{\frac{8\pi G}{3}}\ .
\end{equation}
The present value of $a_{D}$ was set equal to unity.
According to (\ref{rhoD}) the energy density changes from $\rho_{D}\propto a_{D}^{-3}$ at $a_{D}\ll 1$ to
to an approximately constant $\rho_{D}$ at $a_{D}\gg 1$. In the distant past it behaves like matter, in the distant future it mimics a cosmological constant.  It is this feature that made viscous fluids candidates for a unified description of dark matter and dark energy.

The deceleration parameter is given by
\begin{equation}
q_{D} = - 1 - \frac{a_{D}\mathcal{H}_{D}^{\prime}}{\mathcal{H}_{D}} = \frac{-1 + \frac{1}{2}\left(\frac{\sqrt{\rho_{D0}}}{A_{D}} - 1\right)a_{D}^{-3/2}}{1+ \left(\frac{\sqrt{\rho_{D0}}}{A_{D}} - 1\right)a_{D}^{-3/2}}.
\end{equation}
It is convenient to relate the constant $A_{D}$ to the present value $q_{D0}$ of $q_{D}$ :
\begin{equation}
\label{Aq}
A_{D} = \frac{1}{3}\sqrt{\rho_{D0}}\left(1 - 2 q_{D0}\right).
\end{equation}
For  $\zeta_{D}= 0$ one has $A_{D}=0$ and, consequently, $q_{D0} = \frac{1}{2}$ which is the correct limit of an Einstein-de Sitter universe.

Introducing the abbreviations
\begin{equation}
Q_{1} \equiv 1 + q_{D0}\ , \qquad Q_{2} \equiv 1 - 2q_{D0},
\end{equation}
the Hubble rate in terms of the present value of the deceleration parameter then becomes
\begin{equation}\label{HDQ}
\mathcal{H}_{D} = \frac{1}{3}\mathcal{H}_{D0}\left[Q_{2} + 2Q_{1}a_{D}^{-3/2}\right].
\end{equation}
Since $\left\langle\rho_{m}\right\rangle_{D} =
\left\langle\rho_{m}\right\rangle_{D0}a_{D}^{-3}$, the energy density that corresponds to the viscous fluid is
the difference between the total and the matter energies,
\begin{equation}\label{rhoV}
\rho_{vD} = \rho_{D} - \left\langle\rho_{m}\right\rangle_{D0}a_{D}^{-3}
= \frac{1}{9}\,\rho_{D0}\,\left[Q_{2} + 2 Q_{1}a_{D}^{-\frac{3}{2}}\right]^{2}
- \left\langle\rho_{m}\right\rangle_{D0}a_{D}^{-3}
\end{equation}
and for the pressure we have
\begin{equation}\label{pV}
p_{vD} = - \frac{1}{9}\,\rho_{D0}\,Q_{2}\left[Q_{2} + 2 Q_{1}a_{D}^{-3/2}\right].
\end{equation}
From  (\ref{HDQ}) -  (\ref{pV}) it is obvious that the dynamics is entirely determined by the parameters
$\mathcal{H}_{D0}$, $q_{D0}$ and $\left\langle\rho_{m}\right\rangle_{D0}$.

\section{Viscous fluid from backreaction?}
\label{vback}

Let us investigate now, whether backreaction
and the averaged scalar curvature in the domain $D$ can be modeled in terms of an effective bulk viscosity. We identify the energy density in  (\ref{rhob}) with (\ref{rhoV}) and the pressure in (\ref{rhob}) with  (\ref{pV}), i.e.,
\begin{equation}\label{Q+W}
\rho_{vD} = - \frac{1}{16\pi G}\left(\mathcal{Q}_{D} + \mathcal{R}_{D}\right) = \frac{1}{9}\,\rho_{D0}\,\left[Q_{2} + 2 Q_{1}a_{D}^{-\frac{3}{2}}\right]^{2} - \left\langle\rho_{m}\right\rangle_{D0}a_{D}^{-3}
\end{equation}
and
\begin{equation}\label{Q-W/3}
p_{vD} = - \frac{1}{16\pi G}\left(\mathcal{Q}_{D} - \frac{\mathcal{R}_{D}}{3}\right)
= - \frac{1}{9}\,\rho_{D0}\,Q_{2}\left[Q_{2} +2Q_{1}a_{D}^{-3/2}\right]\ .
\end{equation}
Solving for the backreaction quantities $\mathcal{Q}_{D}$ and $\mathcal{R}_{D}$ results in
\begin{equation}
\mathcal{Q}_{D} = - 4\pi G \left(\rho_{vD} + 3 p_{vD}\right)\quad \mathrm{and} \quad  \mathcal{R}_{D} = -12 \pi G \left(\rho_{vD} - p_{vD}\right),
\end{equation}
respectively.
The combination
\begin{equation}
\mathcal{R}_{D} - 3 \mathcal{Q}_{D} = 48\pi G p_{vD}\
\end{equation}
determines the effective pressure.
Further, we have
\begin{equation}\label{rho+3p}
\rho_{D} + 3 p_{D} = \left\langle \rho_{m}\right\rangle_{D} - \frac{1}{4\pi G}\mathcal{Q}_{D}\ .
\end{equation}
From relation (\ref{rho+3p})
it follows that accelerated expansion requires $\mathcal{Q}_{D} > 4\pi G\left\langle \rho_{m}\right\rangle_{D}$.

\noindent
We may relate the present value of the deceleration parameter to the present
values  $\mathcal{Q}_{D0}$ and $\mathcal{R}_{D0}$:
\begin{equation}
q_{D0} = \frac{1}{2}\frac{1 - \frac{\mathcal{Q}_{D0}}{4\pi G \left\langle\rho_{m}\right\rangle_{D0}}}
{1 - \frac{\mathcal{R}_{D0} + \mathcal{Q}_{D0}}{16\pi G \left\langle\rho_{m}\right\rangle_{D0}}}\ .
\end{equation}
Obviously, for vanishing $\mathcal{Q}_{D0}$ and $\mathcal{R}_{D0}$ we recover $q_{D0} = \frac{1}{2}$.
Since positivity of $\rho_{vD}$ implies $\mathcal{R}_{D0} + \mathcal{Q}_{D0} < 0$, we reproduce $\mathcal{Q}_{D0} > 4\pi G \left\langle\rho_{_{m}}\right\rangle_{D0}$ to have $q_{D0} <0$, i.e., accelerated expansion at the present time.
The backreaction has to be larger than a certain threshold value.

Combining (\ref{defA}) and (\ref{Aq}), we find an explicit expression for the bulk-viscosity coefficient in terms of the present values of backreaction and averaged curvature:
\begin{equation}
\label{zetaD}
\zeta_{D} = \frac{1}{9}\frac{\left\langle\rho_{m}\right\rangle_{D0}}{\mathcal{H}_{D0}}
\frac{\mathcal{R}_{D0} - 3 \mathcal{Q}_{D0}}{\mathcal{R}_{D0} + \mathcal{Q}_{D0} - 16\pi G \left\langle\rho_{m}\right\rangle_{D0}}
= \frac{1}{9}\frac{\left\langle\rho_{m}\right\rangle_{D0}}{\mathcal{H}_{D0}}\left(1 - 2q_{D0}\right).
\end{equation}
For $\mathcal{R}_{D0} + \mathcal{Q}_{D0} < 0$, as required for a positive energy, the denominator is always negative.
For any positive $\mathcal{Q}_{D0}$ the numerator is negative as well.
Recall that $\mathcal{R}_{D0} - 3 \mathcal{Q}_{D0} = 48\pi G p_{v0}^{D}$.
Under these conditions the bulk-viscosity coefficient is always positive.
Notice, however, that the effective bulk viscous energy is something artificial and there is no reason why it should always be positive.
The second equation (\ref{zetaD}) relates $\zeta_{D}$ directly to the present value of the deceleration parameter $q_{D0}$. Consistently,
$q_{D0} = \frac{1}{2}$ corresponds to $\zeta_{D}=0$.

\begin{figure}[h!]
{
\includegraphics[width=10cm]{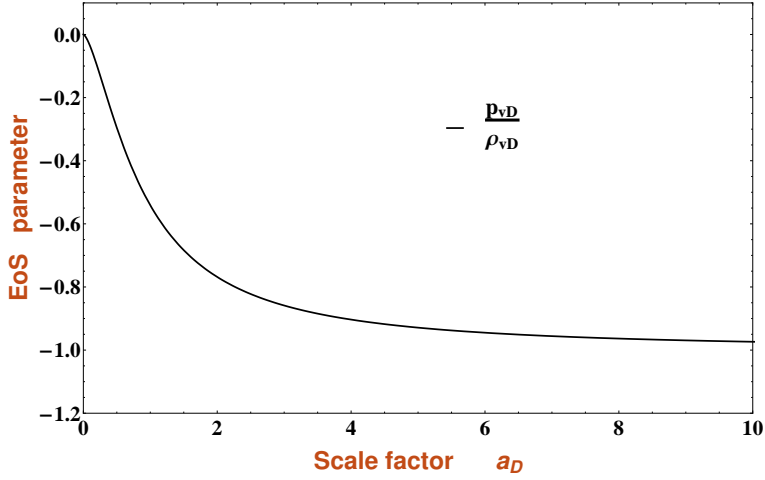}
}
\caption{EoS parameter of the backreaction fluid in terms of the volume scale factor.}
\label{figeos}
\end{figure}

\begin{figure}[h!]
{
\includegraphics[width=10cm]{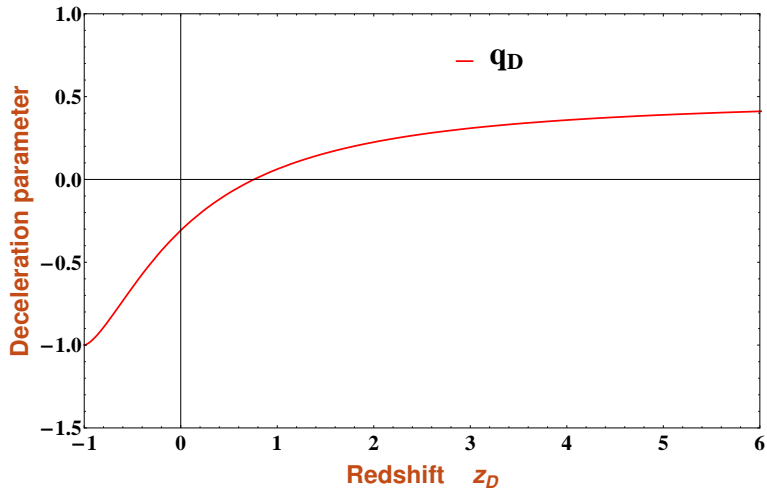}
}
\caption{Dependence of the deceleration parameter $q_{D}$ on the redshift parameter
$z_{D} = a_{D}^{-1} - 1$.}
\label{figdec}
\end{figure}

\begin{figure}[h!]
{
\includegraphics[width=10cm]{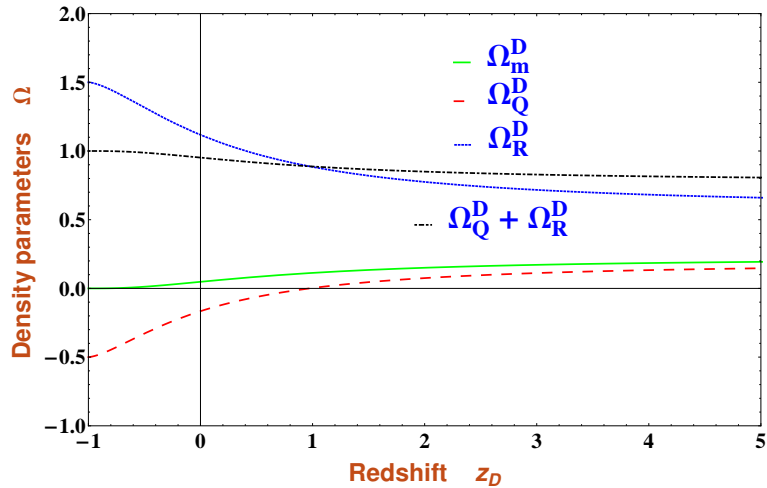}
}
\caption{Dependence of the fractional abundances on the redshift parameter
$z_{D} = a_{D}^{-1} - 1$.}
\label{figomega}
\end{figure}

For $a_{D}\ll 1$ the $a^{-3}_{D}$ terms in (\ref{Q+W}) and (\ref{Q-W/3}) are dominating the backreaction energy density,
\begin{eqnarray}\label{rhob<}
\rho_{vD}(a_{D}\ll 1) & = & \left[\frac{4}{9}\left(q_{D0}- \frac{1}{2}\right)\left(q_{D0}+ \frac{5}{2}\right) \left\langle\rho\right\rangle_{D0}\right. \nonumber\\
& & \left.\quad +
\left(1 + \frac{4}{9}\left(q_{D0}- \frac{1}{2}\right)\left(q_{D0}+ \frac{5}{2}\right)\right)\left\langle\rho_{m}\right\rangle_{D0}\right]a_{D}^{-3}.
\end{eqnarray}
It behaves as non-relativistic matter.
Correspondingly, the effective EoS parameter (\ref{EoS}) tends to zero for $a_{D}\ll 1$.
In the opposite limit $a_{D}\gg 1$,
\begin{equation}\label{pv>}
p_{vD} = - \rho_{vD}= - \frac{1}{9}\rho_{D0}\left(1 - 2q_{0}\right)^{2}\ \qquad (a_{D}\gg 1),
\end{equation}
it acts as a cosmological constant.
The volume-scale-factor dependence of the effective EoS parameter of the backreaction fluid, $\frac{p_{vD}}{\rho_{vD}}$, is shown in Fig.~\ref{figeos} for the best-fit values given in the first entry in Table~\ref{table}.
Explicitly, $\mathcal{Q}_{D}$ and $\mathcal{R}_{D}$ become
\begin{equation}\label{QDa}
\mathcal{Q}_{D}
= \frac{1}{3} \mathcal{H}_{D0}^{2}\,\left[\left(Q_{2} - Q_{1}a_{D}^{-3/2}\right)\left(Q_{2} + 2Q_{1}a_{D}^{-3/2}\right)\right]
+ \frac{3}{2}\mathcal{H}_{D0}^{2}\Omega_{m0}^{D}a_{D}^{-3}\
\end{equation}
and
\begin{equation}\label{RDa}
\mathcal{R}_{D}
= -  \mathcal{H}_{D0}^{2}\,\left[\left(Q_{2} + Q_{1}a_{D}^{-3/2}\right)\left(Q_{2} + 2Q_{1}a_{D}^{-3/2}\right)\right]
+ \frac{9}{2}\mathcal{H}_{D0}^{2}\Omega_{m0}^{D}a_{D}^{-3},
\end{equation}
respectively, where $\Omega_{m0}^{D} = \frac{\left\langle\rho_{m}\right\rangle_{D0}}{\rho_{D0}}$. The corresponding fractional quantities are
\begin{equation}\label{OM}
\Omega_{m}^{D} = \frac{9\,\Omega_{m0}^{D} a_{D}^{-3}}{\left[Q_{2} + 2Q_{1}a_{D}^{-3/2}\right]^{2}},
\qquad
\Omega_{Q}^{D} = -\frac{1}{2}\frac{Q_{2} - Q_{1}a_{D}^{-3/2}}{Q_{2} + 2Q_{1}a_{D}^{-3/2}} - \frac{1}{4}\Omega_{m}^{D},
\end{equation}
and
\begin{equation}\label{OR}
\Omega_{R}^{D} = \frac{3}{2}\frac{Q_{2} + Q_{1}a_{D}^{-3/2}}{Q_{2} + 2Q_{1}a_{D}^{-3/2}} - \frac{3}{4}\Omega_{m}^{D}.
\end{equation}
The resulting deceleration parameter (\ref{qDO}) is shown in Fig.~\ref{figdec}.
For the present values of $\Omega_{Q}^{D}$  and $\Omega_{Q}^{D}$ we have
\begin{equation}\label{OM0}
\Omega_{Q0}^{D} = -\frac{1}{4}\left(1 - 2q_{D0}\right) + \frac{1}{4}\left(1 - \Omega_{m0}^{D}\right)\ \mathrm{and} \ \
\Omega_{R0}^{D} = \frac{1}{4}\left(1 - 2q_{D0}\right) + \frac{3}{4}\left(1 - \Omega_{m0}^{D}\right),
\end{equation}
respectively.
The Einstein-de Sitter universe is recovered for $q_{D0} = \frac{1}{2}$ with $\Omega_{m0}^{D} = 1$, corresponding
to $\Omega_{Q0}^{D}=\Omega_{R0}^{D} = 0$.
Notice that, since $\Omega_{m0}^{D} \leq 1$, the fractional kinematic backreaction $\Omega_{Q0}^{D}$ becomes negative for
$q_{D0}<0$. This means, $\mathcal{Q}_{D}$ itself becomes positive, a necessary ingredient for accelerated expansion
according to (\ref{rho+3p}).
The joint contribution $\Omega_{Q0}^{D}+\Omega_{R0}^{D}$, however, remains positive.
In the high-redshift limit $a_{D}\ll 1$ the fractional abundances take the values
\begin{equation}\label{}
\Omega_{m}^{D} = \frac{9}{4}\frac{\Omega_{m0}^{D}}{\left(1 +q_{D0}\right)^{2}},\quad
\Omega_{Q}^{D} = \frac{1}{4}\left(1 - \Omega_{m}^{D}\right),\quad
\Omega_{R}^{D} = \frac{3}{4}\left(1 - \Omega_{m}^{D}\right)\quad (a_{D}\ll 1).
\end{equation}
For $\Omega_{m0}^{D} < \frac{4}{9}\left(1 +q_{D0}\right)^{2}$ one has $\Omega_{m}^{D} < 1$ and both $\Omega_{Q}^{D}$ and $\Omega_{R}^{D}$ are positive.
For a present deceleration parameter of the order of $q_{D0}\approx - \frac{1}{2}$, the value of $\Omega_{m}^{D}$ at
$a_{D}\ll 1$ is about one order of magnitude larger than $\Omega_{m0}^{D}$. One has to keep in mind, however, that the backreaction fluid itself behaves as matter in this limit.
This restriction on $\Omega_{m0}^{D}$ is compatible with the circumstance that it describes baryonic matter exclusively.
From the outset there is no separate dark-matter component here since the viscous fluid is supposed to account for the entire dark sector.
But in fact the condition $\Omega_{m0}^{D} < \frac{4}{9}\left(1 +q_{D0}\right)^{2}$ leaves room for a matter abundance
somewhat larger than that attributed to the baryons with a fraction of the order of $0.048$.
In Fig.~\ref{figomega} the behavior of the fractional abundances is visualized for the best-fit values of Table~\ref{table}, assuming $\Omega_{m0}^{D}$ to describe the baryonic matter fraction with $\Omega_{m0}^{D} = 0.048$.

\section{Effective metric and luminosity distance}
\label{effective}

\begin{figure}[h!]
{
\includegraphics[width=10cm]{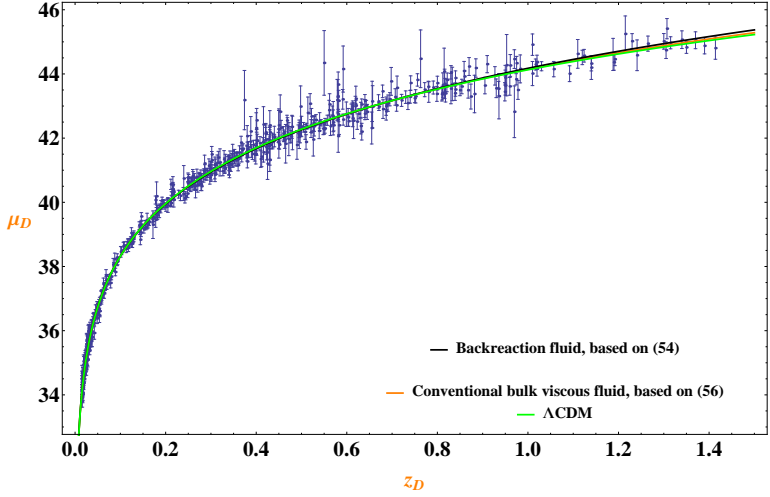}
}
\caption{Dependence of the distance modulus on the redshift parameter
$z_{D} = a_{D}^{-1} - 1$.}
\label{figmu}
\end{figure}

\begin{figure}[h!]
{
\includegraphics[width=10cm]{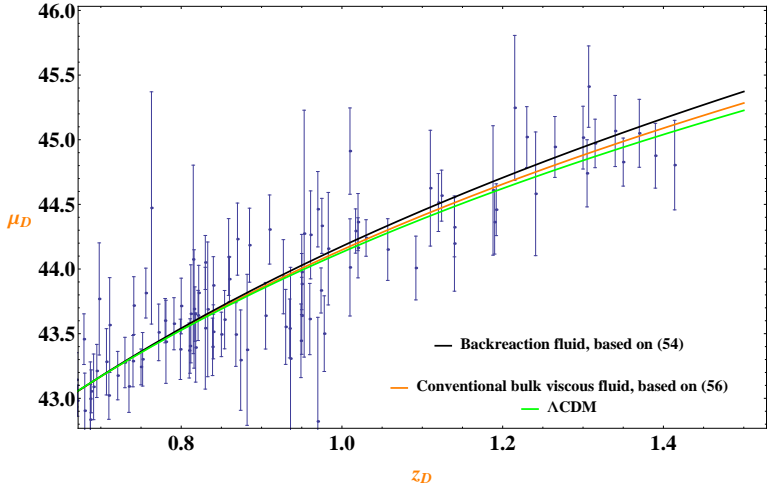}
}
\caption{Magnified region of FIG.~\ref{figmu}.}
\label{figampl}
\end{figure}

\begin{figure}[h!]
{
\includegraphics[width=10cm]{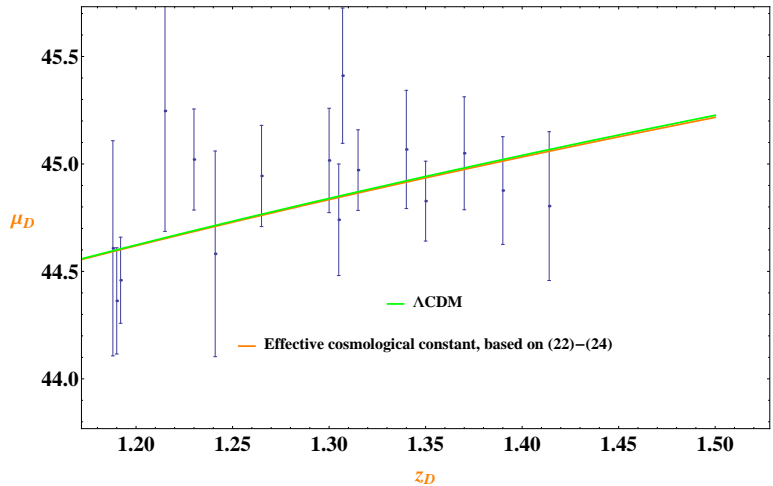}
}
\caption{Dependence of the distance modulus on the redshift parameter for the $\Lambda$CDM model and its backreaction counterpart}
\label{figlcdm}
\end{figure}

As already mentioned, to make contact with observations we have to rely on an additional ingredient which is not part of the formalism so far. Although we emphasized that the definition of the volume scale factor $a_{D}$ is not related to a space-time metric, we shall follow here the practice in the literature \cite{ParanjapeSingh:2006,Larena:2008be} and assume the existence of an effective metric in which $a_{D}$ now indeed mimics a scale factor of a RW type metric which, however, is not required to satisfy the field equations. This effective metric has also to account for the averaged curvature $\mathcal{R}_{D}$ for which we found the expression (\ref{RDa}) and the corresponding fractional quantity (\ref{OR}).
These quantities define a curvature radius
\begin{equation}\label{RcD2}
\mathcal{R}_{cD} = \frac{c}{a_{D}}\sqrt{\frac{-6}{\mathcal{R}_{D}}}
= \frac{c}{a_{D}\mathcal{H}_{D}}\frac{1}{\sqrt{\Omega_{R}^{D}}},
\end{equation}
the present value of which is of the order of the domain-dependent present Hubble radius.
This suggests an effective metric with the effective scale factor $a_{D}$ (cf. Roukema et al. \cite{Roukema:2013}),
\begin{equation}\label{metriccurv}
ds^{2}_{\mathrm{eff}} = - c^{2}dt^{2} + a_{D}^{2}\left[dr^{2} + \mathcal{R}_{cD}^{2}\sinh^{2} \frac{r}{\mathcal{R}_{cD}}\left(d\vartheta^{2}
+ \sin^{2}\vartheta d\varphi^{2}\right)\right],
\end{equation}
generalizing the RW metric which is recovered for $\mathcal{R}_{cD}^{-1} = \sqrt{|k|} = \ $ const. Here, $\mathcal{R}_{cD}$ is a time-dependent quantity such that each slice $t=\ $ constant is characterized by a different curvature. By comparing the backreaction dynamics with the dynamics based  on the standard RW metric we
assume an averaging volume of the size of the observable Universe.
On this basis radial light propagation is described by
\begin{equation}\label{}
ds^{2}_{\mathrm{eff}} = 0 \quad \Rightarrow\quad dr = \frac{c}{a_{D}^{2}\mathcal{H}_{D}}da_{D}.
\end{equation}
With (\ref{HDQ}) we find for $r(a_{D})$,
\begin{eqnarray}\label{r(a)}
r(a_{D}) &=& -\frac{6c}{\mathcal{H}_{D0}}\frac{1}{6\sqrt[3]{Q_{2}}\,(2Q_{1})^{2/3}}
\left\{\ln \left[\frac{Q_{2}^{2/3}a_{D} -\sqrt[3]{Q_{2}}\,\sqrt[3]{2Q_{1}}\,a_{D}^{1/2} + (2Q_{1})^{2/3}}{Q_{2}^{2/3} -\sqrt[3]{Q_{2}}\,\sqrt[3]{2Q_{1}}\, + (2Q_{1})^{2/3}}\right]\right.\nonumber\\
&&\left. - 2\ln\left[\frac{\sqrt[3]{Q_{2}}\,a_{D}^{1/2} + \sqrt[3]{2Q_{1}}}{\sqrt[3]{Q_{2}}\, + \sqrt[3]{2Q_{1}}}\right]\right.\nonumber\\
&&\left. + 2 \sqrt{3}\left[\arctan\frac{1 - 2\frac{\sqrt[3]{Q_{2}}}{\sqrt[3]{2Q_{1}}}\,a_{D}^{1/2}}{\sqrt{3}}
- \arctan\frac{1 - 2\frac{\sqrt[3]{Q_{2}}}{\sqrt[3]{2Q_{1}}}}{\sqrt{3}}
\right]
\right\}.
\end{eqnarray}
Introducing an effective redshift prameter $z_{D}$ by $1 + z_{D} = a_{D}^{-1}$,
we may calculate the luminosity distance $d_{L}^{\mathrm{eff}}(z_{D})$ by
\begin{equation}\label{dLcurv}
d_{L}^{\mathrm{eff}}(z_{D}) = \left(1+z_{D}\right)\mathcal{R}_{cD}(z_{D})\sinh \frac{r(z_{D})}{\mathcal{R}_{cD}(z_{D})}
\end{equation}
and determine a distance modulus
\begin{equation}
\mu_{D}=5\log d_L^{\mathrm{eff}}(z_{D}) +\mu_{D0}
\label{moduloD}
\end{equation}
with $\mu_{D0}=42.384-5\log h_{D}$, where $h_{D}$ is defined by $\mathcal{H}_{D0} = 100  h_{D} \mathrm{km s^{-1} Mpc^{-1}}$. By adopting the values from the standard analysis we imply that the averaging scale is the size of the observable Universe.
 Then we may contrast these relations with those of a spatially flat ``conventional" bulk viscous fluid universe where the bulk viscous fluid is part of the matter sector.
The formal calculations then are very similar to the steps in section \ref{vfluid}. In particular, the Hubble rate has the structure (\ref{HDQ}) as well. The difference is that the volume scale factor $a_{D}$ in (\ref{HDQ}) is replaced by the scale factor $a$ of the RW metric and the subscripts D become superfluous.
Then $H(z) = \frac{1}{3}H_{0}\left[1 - 2 q_{0} + 2\left(1 + q_{0}\right)(1+z)^{3/2}\right]$.
For such a model the luminosity distance is obtained via the standard formula
\begin{equation}
d_{L}=\left(z+1\right) c \int_{0}^{z}\frac{dz'}{H\left(z'\right)},
\label{luminosidade}
\end{equation}
which is used in the distance modulus
\begin{equation}
\mu=5\log d_L(z)+\mu_0 \
\label{modulo}
\end{equation}
with $\mu_0=42.384-5\log h $.
In Figs.~\ref{figmu} and \ref{figampl}, using the data from the Union2.1 sample \cite{suzuki},  we compare the results for the backreaction fluid model based on (\ref{dLcurv}) with those of the
``conventional" spatially flat bulk viscous model based on (\ref{luminosidade}). While there are differences, these are apparently not large enough to clearly discriminate between both models. For comparison we have also included the result for the $\Lambda$CDM model. As far as the analysis of the supernova data is concerned, our conclusion is that  the backreaction model with its large curvature contribution yields similar results as the corresponding flat model.
This brings us also back to the $\Lambda$CDM dynamics with a backreaction induced effective cosmological constant. Either the luminosity distance can be calculated via the standard formula (\ref{luminosidade}) with the standard Hubble rate of the $\Lambda$CDM universe,
$H^{\Lambda CDM}(z) = H^{\Lambda CDM}_{0}\sqrt{\Omega_{m0}(1+z)^{3} + \Omega_{\Lambda}}$
 or, in the backreaction context
via (\ref{dLcurv}) and the domain-dependent Hubble function $\mathcal{H}_{D}^{\Lambda CDM}(z) = \mathcal{H}^{\Lambda CDM}_{D0}\sqrt{\Omega^{D}_{m0}(1+z)^{3} + \Omega^{D}_{\Lambda}}$
with the cosmological backreaction constant (\ref{LambdaD}) and the curvature radius (\ref{RcDL}).
Using the SNIa sample again, we look for the best-fit values for each of these cases which are shown in the last two entries in Table~\ref{table}.
The results are presented in Fig.~\ref{figlcdm}.
The standard $\Lambda$CDM model seems to fare slightly better but without being clearly superior to the backreaction dynamics.

\begin{table}[h!] \centering
\caption{Results of the statistical analysis based on the Union2.1 data.}
\begin{tabular}{|c||c||c||c|}
\hline
\multicolumn{4}{|c|}{
SN Ia Union 2.1} \\ \hline \hline
& & & \\
$Backreaction~~fluid~~for~~\Omega_{m0}=0.048$  &  $\chi^{2}_{\nu}$  & $q_0$ (95\% CL) & $h$   \\
& & & \\ \hline
& & & \\
                  &         $0.985$        & $-0.307_{-0.073}^{+0.070}$  & $0.693$   \\
& & & \\ \hline
& & & \\
$Conventional~~bulk~~viscous~~fluid$  &  $\chi^{2}_{\nu}$  & $q_0$ (95\% CL)  & $h$   \\
& & & \\ \hline
& & & \\
                  &         $0.974$        & $-0.480_{-0.059}^{+0.066}$  & $0.697$   \\
& & & \\ \hline
& & & \\
Effective cosmological constant &     $\chi^{2}_{\nu}$    &     $\Omega_{m0}$ (95\% CL)      &     $h$     \\
& & & \\ \hline
& & & \\
                  &         $0.971$        & $0.318^{+0.051}_{-0.041}$  & $0.699$   \\
& & & \\ \hline
& & & \\
$\Lambda$CDM &     $\chi^{2}_{\nu}$    &     $\Omega_{m0}$ (95\% CL)      &     $h$     \\
& & & \\ \hline
& & & \\
                  &         $0.971$        & $0.279^{+0.038}_{-0.039}$  & $0.701$   \\
& & & \\ \hline

\end{tabular}
\label{table}
\end{table}

\section{Summary}
\label{summary}

We investigated the possibility that a cosmological bulk viscous fluid dynamics is the result of a backreaction due to averaged inhomogeneities in a pure dust universe. We quantified the dynamical backreaction and the averaged curvature throughout the matter period of the cosmological evolution. With the help of an effective metric with time-dependent spatial curvature we tested this model against the SNIa observations of the Union2.1 sample.
We did not find substantial differences to the results of a standard analysis for a spatially flat bulk-viscous cosmology.
To the best of our knowledge such type of interpreting observations of one and the same specific model both in the backreaction context and in the standard way has not been performed so far.
A similar comment holds for an effective cosmological constant model in which the dynamics of a cosmological constant is mimicked by a combination of kinematic backreaction and average curvature.
In this perspective the backreaction picture is apparently compatible with respect to the interpretation of the SNIa data.
Whether or not it will pass also other tests will be the subject of future research.

The crucial difference between the backreaction models and the standard FLRW descriptions is the existence of a curvature evolution in the former while the FLRW models come with a constant (in most cases vanishing) curvature.
In \cite{clarkson} a useful quantity has been introduced which is identically zero for all FLRW configurations but not for more general models. For a class of
curvature-evolution models this quantity was discussed in \cite{Larena:2008be}. On this basis, according to the authors of \cite{Larena:2008be},
future results from the Euclid mission \cite{euclid} are expected to be able to discriminate between curvature-evolution models
and models with constant curvature.


%


\begin{acknowledgements}
We are grateful to the anonymous  referee for constructive comments.
We thank CAPES, FAPES and CNPq (Brazil) for financial support.
\end{acknowledgements}



\end{document}